\RequirePackage{lineno} 
\documentclass[12pt]{iopart}

\usepackage{feynmf}

\usepackage{graphicx,epsfig}
\usepackage{longtable}
\usepackage{color}
\usepackage{footmisc}
\usepackage[sort&compress,numbers]{natbib}
\usepackage{doi}
\usepackage{hyperref}

\usepackage{bm}
\usepackage{capt-of}
\usepackage{relsize}
\usepackage{xcolor}
\usepackage{amssymb}
\usepackage[normalem]{ulem}
\usepackage{mathptmx}

\def\al{\alpha}

\def\ga{\gamma}
\def\de{\delta}

\def\la{\lambda}

\def\si{\sigma}

\def\ph{\phi}

\def\om{\omega}

\def\De{\Delta}

\def\nue{\nu_e}
\def\numu{\nu_\mu}

\def\nubar{\bar\nu}
\def\nuebar{\bar\nu_e}
\def\numubar{\bar\nu_\mu}

\newcommand{\beq}{\begin{eqnarray}}
\newcommand{\eeq}{\end{eqnarray}}
\def\to{\rightarrow}

\def\piz{\pi^0}
\def\EnuAve{\left<E_\nu\right>}

\def\POT{5.738\times 10^{20}}
\def\T2KEnu{0.6} 
\def\magnet{0.2}
\def\dpipe{96}
\def\target{5.54\times 10^{29}}
\def\vNEUT{5.3.2}
\def\recentnue{89}
\def\recentnuebar{7}

\def\NOMADEnu{25}

\def\NOMADratioA{4.0\times 10^{-4}}
\def\NOMADlimA{0.0068\times 10^{-38}}

\def\purity{95}     %
\def\efficiency{1.9}
\def\Nevt{46}
\def\NevtPhi{39}

\def\PhiL{252^\circ}
\def\PhiH{288^\circ}

\def\FidX{174.9}
\def\FidY{174.9}
\def\FidZ{54.2} 

\def\psta{14.7} 
\def\psyb{26.8}\def\nsyb{16.0} 
\def\psyp{15.4}\def\nsyp{13.4} 
\def\psyx{4.3}\def\nsyx{3.8} 
\def\psyf{7.7} 
\def\psyd{6.6} 
\def\ptot{30.6}\def\ntot{21.0} 

\def\CCincla{6} 
\def\CCinclb{38}
\def\eFGDmass{0.6}

\def\limA{0.114\times 10^{-38}} 

\def\EMFP{\lambda_{EMFP}}

\newcommand{\ININSTHD}{\address{$^{1} $University Autonoma Madrid, Department of Theoretical Physics, 28049 Madrid, Spain}}
\newcommand{\ININSTEE}{\address{$^{2} $University of Bern, Albert Einstein Center for Fundamental Physics, Laboratory for High Energy Physics (LHEP), Bern, Switzerland}}
\newcommand{\ININSTFE}{\address{$^{3} $Boston University, Department of Physics, Boston, Massachusetts, U.S.A.}}
\newcommand{\ININSTD} {\address{$^{4} $University of British Columbia, Department of Physics and Astronomy, Vancouver, British Columbia, Canada}}
\newcommand{\ININSTGA}{\address{$^{5} $University of California, Irvine, Department of Physics and Astronomy, Irvine, California, U.S.A.}}
\newcommand{\ININSTI} {\address{$^{6} $IRFU, CEA Saclay, Gif-sur-Yvette, France}}
\newcommand{\ININSTGB}{\address{$^{7} $University of Colorado at Boulder, Department of Physics, Boulder, Colorado, U.S.A.}}
\newcommand{\ININSTFG}{\address{$^{8} $Colorado State University, Department of Physics, Fort Collins, Colorado, U.S.A.}}
\newcommand{\ININSTFH}{\address{$^{9} $Duke University, Department of Physics, Durham, North Carolina, U.S.A.}}
\newcommand{\ININSTBA}{\address{$^{10}$Ecole Polytechnique, IN2P3-CNRS, Laboratoire Leprince-Ringuet, Palaiseau, France }}
\newcommand{\ININSTEF}{\address{$^{11}$ETH Zurich, Institute for Particle Physics, Zurich, Switzerland}}
\newcommand{\ININSTEG}{\address{$^{12}$University of Geneva, Section de Physique, DPNC, Geneva, Switzerland}}
\newcommand{\ININSTDG}{\address{$^{13}$H. Niewodniczanski Institute of Nuclear Physics PAN, Cracow, Poland}}
\newcommand{\ININSTCB}{\address{$^{14}$High Energy Accelerator Research Organization (KEK), Tsukuba, Ibaraki, Japan}}
\newcommand{\ININSTIB}{\address{$^{15}$University of Houston, Department of Physics, Houston, Texas, U.S.A.}}
\newcommand{\ININSTED}{\address{$^{16}$Institut de Fisica d'Altes Energies (IFAE), The Barcelona Institute of Science and Technology, Campus UAB, Bellaterra (Barcelona) Spain}}
\newcommand{\ININSTEC}{\address{$^{17}$IFIC (CSIC \& University of Valencia), Valencia, Spain}}
\newcommand{\ININSTHH}{\address{$^{18}$Institute For Interdisciplinary Research in Science and Education (IFIRSE), ICISE, Quy Nhon, Vietnam}}
\newcommand{\ININSTEI}{\address{$^{19}$Imperial College London, Department of Physics, London, United Kingdom}}
\newcommand{\ININSTGF}{\address{$^{20}$INFN Sezione di Bari and Universit\`a e Politecnico di Bari, Dipartimento Interuniversitario di Fisica, Bari, Italy}}
\newcommand{\ININSTBE}{\address{$^{21}$INFN Sezione di Napoli and Universit\`a di Napoli, Dipartimento di Fisica, Napoli, Italy}}
\newcommand{\ININSTBF}{\address{$^{22}$INFN Sezione di Padova and Universit\`a di Padova, Dipartimento di Fisica, Padova, Italy}}
\newcommand{\ININSTBD}{\address{$^{23}$INFN Sezione di Roma and Universit\`a di Roma ``La Sapienza'', Roma, Italy}}
\newcommand{\ININSTEB}{\address{$^{24}$Institute for Nuclear Research of the Russian Academy of Sciences, Moscow, Russia}}
\newcommand{\ININSTHI}{\address{$^{25}$Institute of Physics (IOP), Vietnam Academy of Science and Technology (VAST), Hanoi, Vietnam}}
\newcommand{\ININSTHA}{\address{$^{26}$Kavli Institute for the Physics and Mathematics of the Universe (WPI), The University of Tokyo Institutes for Advanced Study, University of Tokyo, Kashiwa, Chiba, Japan}}
\newcommand{\ININSTCC}{\address{$^{27}$Kobe University, Kobe, Japan}}
\newcommand{\ININSTCD}{\address{$^{28}$Kyoto University, Department of Physics, Kyoto, Japan}}
\newcommand{\ININSTEJ}{\address{$^{29}$Lancaster University, Physics Department, Lancaster, United Kingdom}}
\newcommand{\ININSTFC}{\address{$^{30}$University of Liverpool, Department of Physics, Liverpool, United Kingdom}}
\newcommand{\ININSTFI}{\address{$^{31}$Louisiana State University, Department of Physics and Astronomy, Baton Rouge, Louisiana, U.S.A.}}
\newcommand{\ININSTHB}{\address{$^{32}$Michigan State University, Department of Physics and Astronomy,  East Lansing, Michigan, U.S.A.}}
\newcommand{\ININSTCE}{\address{$^{33}$Miyagi University of Education, Department of Physics, Sendai, Japan}}
\newcommand{\ININSTDF}{\address{$^{34}$National Centre for Nuclear Research, Warsaw, Poland}}
\newcommand{\ININSTFJ}{\address{$^{35}$State University of New York at Stony Brook, Department of Physics and Astronomy, Stony Brook, New York, U.S.A.}}
\newcommand{\ININSTGJ}{\address{$^{36}$Okayama University, Department of Physics, Okayama, Japan}}
\newcommand{\ININSTCF}{\address{$^{37}$Osaka City University, Department of Physics, Osaka, Japan}}
\newcommand{\ININSTGG}{\address{$^{38}$Oxford University, Department of Physics, Oxford, United Kingdom}}
\newcommand{\ININSTGC}{\address{$^{39}$University of Pittsburgh, Department of Physics and Astronomy, Pittsburgh, Pennsylvania, U.S.A.}}
\newcommand{\ININSTFA}{\address{$^{40}$Queen Mary University of London, School of Physics and Astronomy, London, United Kingdom}}
\newcommand{\ININSTE} {\address{$^{41}$University of Regina, Department of Physics, Regina, Saskatchewan, Canada}}
\newcommand{\ININSTGD}{\address{$^{42}$University of Rochester, Department of Physics and Astronomy, Rochester, New York, U.S.A.}}
\newcommand{\ININSTHC}{\address{$^{43}$Royal Holloway University of London, Department of Physics, Egham, Surrey, United Kingdom}}
\newcommand{\ININSTBC}{\address{$^{44}$RWTH Aachen University, III. Physikalisches Institut, Aachen, Germany}}
\newcommand{\ININSTFB}{\address{$^{45}$University of Sheffield, Department of Physics and Astronomy, Sheffield, United Kingdom}}
\newcommand{\ININSTDI}{\address{$^{46}$University of Silesia, Institute of Physics, Katowice, Poland}}
\newcommand{\ININSTIA}{\address{$^{47}$SLAC National Accelerator Laboratory, Stanford University, Menlo Park, California, USA}}
\newcommand{\ININSTBB}{\address{$^{48}$Sorbonne Universit\'e, Universit\'e Paris Diderot, CNRS/IN2P3, Laboratoire de Physique Nucl\'eaire et de Hautes Energies (LPNHE), Paris, France}}
\newcommand{\ININSTEH}{\address{$^{49}$STFC, Rutherford Appleton Laboratory, Harwell Oxford,  and  Daresbury Laboratory, Warrington, United Kingdom}}
\newcommand{\ININSTCH}{\address{$^{50}$University of Tokyo, Department of Physics, Tokyo, Japan}}
\newcommand{\ININSTBJ}{\address{$^{51}$University of Tokyo, Institute for Cosmic Ray Research, Kamioka Observatory, Kamioka, Japan}}
\newcommand{\ININSTCG}{\address{$^{52}$University of Tokyo, Institute for Cosmic Ray Research, Research Center for Cosmic Neutrinos, Kashiwa, Japan}}
\newcommand{\ININSTHF}{\address{$^{53}$Tokyo Institute of Technology, Department of Physics, Tokyo, Japan}}
\newcommand{\ININSTGI}{\address{$^{54}$Tokyo Metropolitan University, Department of Physics, Tokyo, Japan}}
\newcommand{\ININSTHG}{\address{$^{55}$Tokyo University of Science, Faculty of Science and Technology, Department of Physics, Noda, Chiba, Japan}}
\newcommand{\ININSTF} {\address{$^{56}$University of Toronto, Department of Physics, Toronto, Ontario, Canada}}
\newcommand{\ININSTB} {\address{$^{57}$TRIUMF, Vancouver, British Columbia, Canada}}
\newcommand{\ININSTG} {\address{$^{58}$University of Victoria, Department of Physics and Astronomy, Victoria, British Columbia, Canada}}
\newcommand{\ININSTDJ}{\address{$^{59}$University of Warsaw, Faculty of Physics, Warsaw, Poland}}
\newcommand{\ININSTDH}{\address{$^{60}$Warsaw University of Technology, Institute of Radioelectronics, Warsaw, Poland}}
\newcommand{\ININSTFD}{\address{$^{61}$University of Warwick, Department of Physics, Coventry, United Kingdom}}
\newcommand{\ININSTGH}{\address{$^{62}$University of Winnipeg, Department of Physics, Winnipeg, Manitoba, Canada}}
\newcommand{\ININSTEA}{\address{$^{63}$Wroclaw University, Faculty of Physics and Astronomy, Wroclaw, Poland}}
\newcommand{\ININSTHE}{\address{$^{64}$Yokohama National University, Faculty of Engineering, Yokohama, Japan}}
\newcommand{\ININSTH} {\address{$^{65}$York University, Department of Physics and Astronomy, Toronto, Ontario, Canada}}                                                                                              

\newcommand{\INSTHD}{$^{1} $}
\newcommand{\INSTEE}{$^{2} $}
\newcommand{\INSTFE}{$^{3} $}
\newcommand{\INSTD} {$^{4} $}
\newcommand{\INSTGA}{$^{5} $}
\newcommand{\INSTI} {$^{6} $}
\newcommand{\INSTGB}{$^{7} $}
\newcommand{\INSTFG}{$^{8} $}
\newcommand{\INSTFH}{$^{9} $}
\newcommand{\INSTBA}{$^{10}$}
\newcommand{\INSTEF}{$^{11}$}
\newcommand{\INSTEG}{$^{12}$}
\newcommand{\INSTDG}{$^{13}$}
\newcommand{\INSTCB}{$^{14}$}
\newcommand{\INSTIB}{$^{15}$}
\newcommand{\INSTED}{$^{16}$}
\newcommand{\INSTEC}{$^{17}$}
\newcommand{\INSTHH}{$^{18}$}
\newcommand{\INSTEI}{$^{19}$}
\newcommand{\INSTGF}{$^{20}$}
\newcommand{\INSTBE}{$^{21}$}
\newcommand{\INSTBF}{$^{22}$}
\newcommand{\INSTBD}{$^{23}$}
\newcommand{\INSTEB}{$^{24}$}
\newcommand{\INSTHI}{$^{25}$}
\newcommand{\INSTHA}{$^{26}$}
\newcommand{\INSTCC}{$^{27}$}
\newcommand{\INSTCD}{$^{28}$}
\newcommand{\INSTEJ}{$^{29}$}
\newcommand{\INSTFC}{$^{30}$}
\newcommand{\INSTFI}{$^{31}$}
\newcommand{\INSTHB}{$^{32}$}
\newcommand{\INSTCE}{$^{33}$}
\newcommand{\INSTDF}{$^{34}$}
\newcommand{\INSTFJ}{$^{35}$}
\newcommand{\INSTGJ}{$^{36}$}
\newcommand{\INSTCF}{$^{37}$}
\newcommand{\INSTGG}{$^{38}$}
\newcommand{\INSTGC}{$^{39}$}
\newcommand{\INSTFA}{$^{40}$}
\newcommand{\INSTE} {$^{41}$}
\newcommand{\INSTGD}{$^{42}$}
\newcommand{\INSTHC}{$^{43}$}
\newcommand{\INSTBC}{$^{44}$}
\newcommand{\INSTFB}{$^{45}$}
\newcommand{\INSTDI}{$^{46}$}
\newcommand{\INSTIA}{$^{47}$}
\newcommand{\INSTBB}{$^{48}$}
\newcommand{\INSTEH}{$^{49}$}
\newcommand{\INSTCH}{$^{50}$}
\newcommand{\INSTBJ}{$^{51}$}
\newcommand{\INSTCG}{$^{52}$}
\newcommand{\INSTHF}{$^{53}$}
\newcommand{\INSTGI}{$^{54}$}
\newcommand{\INSTHG}{$^{55}$}
\newcommand{\INSTF} {$^{56}$}
\newcommand{\INSTB} {$^{57}$}
\newcommand{\INSTG} {$^{58}$}
\newcommand{\INSTDJ}{$^{59}$}
\newcommand{\INSTDH}{$^{60}$}
\newcommand{\INSTFD}{$^{61}$}
\newcommand{\INSTGH}{$^{62}$}
\newcommand{\INSTEA}{$^{63}$}
\newcommand{\INSTHE}{$^{64}$}
\newcommand{\INSTH} {$^{65}$}

\begin{document}


\title{Search for neutral-current induced single photon production at the ND280 near detector in T2K}

\author{
  K.\,Abe\INSTBJ
, R.\,Akutsu\INSTCG 
, A.\,Ali\INSTBF
, C.\,Andreopoulos\INSTFC$^,$\INSTEH
, L.\,Anthony\INSTFC
, M.\,Antonova\INSTEC 
, S.\,Aoki\INSTCC
, A.\,Ariga\INSTEE 
, Y.\,Ashida\INSTCD 
, Y.\,Awataguchi\INSTGI
% ...
}
{\color{white}
\footnote[1]{now at CERN}
\footnote[2]{also at J-PARC, Tokai, Japan}
\footnote[3]{affiliated member at Kavli IPMU (WPI), the University of Tokyo, Japan}
\footnote[4]{also at National Research Nuclear University "MEPhI" and Moscow Institute of Physics and Technology, Moscow, Russia}
\footnote[5]{deceased}
\footnote[6]{also at JINR, Dubna, Russia}
\footnote[7]{also at BMCC/CUNY, Science Department, New York, New York, U.S.A.}
}
\hspace{6cm}({\bf The T2K Collaboration})

\ININSTHD
\ININSTEE
\ININSTFE
\ININSTD
\ININSTGA
\ININSTI
\ININSTGB
\ININSTFG
\ININSTFH
\ININSTBA
\ININSTEF
\ININSTEG
\ININSTDG
\ININSTCB
\ININSTIB
\ININSTED
\ININSTEC
\ININSTHH
\ININSTEI
\ININSTGF
\ININSTBE
\ININSTBF
\ININSTBD
\ININSTEB
\ININSTHI
\ININSTHA
\ININSTCC
\ININSTCD
\ININSTEJ
\ININSTFC
\ININSTFI
\ININSTHB
\ININSTCE
\ININSTDF
\ININSTFJ
\ININSTGJ
\ININSTCF
\ININSTGG
\ININSTGC
\ININSTFA
\ININSTE
\ININSTGD
\ININSTHC
\ININSTBC
\ININSTFB
\ININSTDI
\ININSTIA
\ININSTBB
\ININSTEH
\ININSTCH
\ININSTBJ
\ININSTCG
\ININSTHF
\ININSTGI
\ININSTHG
\ININSTF
\ININSTB
\ININSTG
\ININSTDJ
\ININSTDH
\ININSTFD
\ININSTGH
\ININSTEA
\ININSTHE
\ININSTH

\ead{katori@fnal.gov}
\vspace{10pt}
\begin{indented}
\item[]November 2018
\end{indented}

\begin{abstract}
  Neutrino neutral-current induced single photon production is a sub-leading order process for accelerator-based neutrino beam experiments including T2K. It is, however, an important process to understand because it is a background for electron (anti)neutrino appearance oscillation experiments. Here, we performed the first search of this process below 1 GeV using the fine-grained detector at the T2K ND280 off-axis near detector. By reconstructing single photon kinematics from electron-positron pairs, we achieved $\purity$\% pure gamma ray sample from $\POT$ protons-on-targets neutrino mode data. We do not find positive evidence of neutral current induced single photon production in this sample. We set the model-dependent upper limit on the cross-section for this process, at $\limA$~cm$^2$ (90\% C.L.) per nucleon, using the J-PARC off-axis neutrino beam with an average energy of $\EnuAve\sim\T2KEnu$~GeV. This is the first limit on this process below 1~GeV which is important for current and future oscillation experiments looking for electron neutrino appearance oscillation signals.
\end{abstract}

%
\vspace{2pc}

\section{Neutral current single photon production}

Measurements of neutrino oscillation provide an emerging picture of the neutrino Standard Model ($\nu$SM). A series of high precision neutrino oscillation measurements by T2K~\cite{Abe:2011sj,Abe:2013hdq,Abe:2017bay,Abe:2017vif,Abe:2018wpn} and others~\cite{Ahmad:2001an,Ahn:2002up,Eguchi:2002dm,Arpesella:2008mt,Wendell:2010md,Adamson:2014vgd,An:2016ses,Adamson:2017gxd,Aartsen:2017nmd,NOvA:2018gge} are consistent with three massive neutrinos in the Standard Model (SM)~\cite{Esteban:2016qun}. The neutrino oscillation parameters are free parameters in the lepton mixing matrix of the $\nu$SM that are determined from measurements.  Among them, the Dirac CP phase, $\de_{CP}$, is a key parameter to measure since it may shed light on the mystery of matter-antimatter asymmetry of the universe~\cite{Fukugita:1986hr}.  Recently T2K reported the observation of $\recentnue~\nue$ candidate events in $\numu\to\nue$ ($\nue$ appearance), and $\recentnuebar~\nuebar$ candidate events in $\numubar\to\nuebar$ ($\nuebar$ appearance)~\cite{Abe:2018wpn}.  These observations show that CP conserving values, $\de_{CP}=0$ and $\pi$, fall outside the 2$\si$ confidence intervals. Future experiments, Hyper-Kamiokande (Hyper-K)~\cite{Abe:2015zbg} and the Deep Underground Neutrino Experiment~\cite{Acciarri:2016crz} will use higher intensity beams and more massive detectors to make precision measurements of oscillation with $\mathcal{O}(1,000)$ $\nue$ and $\nuebar$ candidate events.  The  $\nue$ and $\nuebar$ appearance channels can also be used to search for unexpected physics processes. The MiniBooNE experiment reports $\nue$($\nuebar$) appearance oscillation candidate signals from $\numu$($\numubar$) dominant beam~\cite{Aguilar-Arevalo:2018gpe}. One interpretation of this event excess is the existence of sterile neutrinos~\cite{Collin:2016rao,Dentler:2018sju}, but the excess may be events from another interaction channel that was not considered.  

Neutral-current (NC) induced photons often contribute to misidentified (misID) backgrounds in $\numu\to\nue$($\numubar\to\nuebar$) oscillation experiments. In these experiments, single isolated electromagnetic showers are signals of $\nue$($\nuebar$) appearance in oscillations from $\numu$ ($\numubar$) dominant beams.  Photons induced by $\numu$($\numubar$) NC interactions could mimic these signal events. There are two relevant backgrounds, NC induced single $\piz$ production (NC1$\piz$)  and NC induced single photon production (NC1$\ga$). NC1$\piz$ can be a significant background if one of two gamma rays fails to be detected. Recent analysis at T2K~\cite{Abe:2017vif} shows that this background can be rejected effectively by introducing a likelihood-based reconstruction technique at the Super-Kamiokande (Super-K) far detector~\cite{Abe:2013hdq}. Similarly, liquid argon time projection chamber (LArTPC) detectors~\cite{Acciarri:2016smi} have achieved comparable photon identification from neutrino-produced $\piz$s~\cite{Acciarri:2015ncl,Adams:2018sgn}.   

\begin{figure}[ht]
  \center
  \begin{fmffile}{directdelta}
    \vspace{2em}
    a)
    \hspace{2em}
    \begin{fmfgraph*}(70,70)
      \fmfstraight
      \fmftop{i2,o3}
      \fmfleft{i1,b2,b3,i2}
      \fmfright{o1,o2,o3}   
      \fmfbottom{i1,o1}
      \fmf{fermion,tension=1.5}{i2,v1}
      \fmflabel{$\nu$}{i2}
      \fmf{fermion}{v1,o3}
      \fmflabel{$\nu$}{o3}
      \fmf{fermion,tension=1.7}{i1,v2}
      \fmflabel{N}{i1}
      \fmf{fermion,label=N,tension=2}{v2,v3}
      \fmf{photon,label=$Z^0$}{v1,v2}
      \fmf{photon,tension=1.5}{v3,o2}
      \fmflabel{$\gamma$}{o2}
      \fmf{fermion}{v3,o1}
      \fmflabel{N}{o1}
    \end{fmfgraph*}
    \hspace{4em}
    b)
    \hspace{2em}
    \begin{fmfgraph*}(70,70)
      \fmfstraight
      \fmftop{i2,o3}
      \fmfleft{i1,b2,b3,i2}
      \fmfright{o1,o2,o3}
      \fmfbottom{i1,o1}
      \fmf{fermion,tension=1.5}{i2,v1}
      \fmflabel{$\nu$}{i2}
      \fmf{fermion}{v1,o3}
      \fmflabel{$\nu$}{o3}
      \fmf{fermion,tension=1.7}{i1,v2}
      \fmflabel{N}{i1}
      \fmf{dbl_plain_arrow,label=N$^*$,tension=2}{v2,v3}
      \fmf{photon,label=$Z^0$}{v1,v2}
      \fmf{photon,tension=1.5}{v3,o2}
      \fmflabel{$\gamma$}{o2}
      \fmf{fermion}{v3,o1}
      \fmflabel{N}{o1}
    \end{fmfgraph*}

    \vspace{4em}
    c)
    \hspace{2em}
    \begin{fmfgraph*}(70,70)
      \fmfstraight
      \fmftop{i2,o4}
      \fmfleft{i1,b2,b3,i2}
      \fmfright{o1,o2,o3}
      \fmfbottom{i1,o1}
      \fmf{fermion}{i2,v1}
      \fmflabel{$\nu$}{i2}
      \fmf{fermion}{v1,o4}
      \fmflabel{$\nu$}{o3}
      \fmf{fermion}{i1,v3}
      \fmflabel{N}{i1}
      \fmf{photon,label=$Z^0$}{v1,v4}
      \fmf{dashes,label=M}{v5,v3}
      \fmf{photon}{v6,o2}
      \fmf{fermion,tension=0.5}{v4,v5,v6,v4}
      \fmflabel{$\gamma$}{o2}
      \fmf{fermion}{v3,o1}
      \fmflabel{N}{o1}
    \end{fmfgraph*}
    \hspace{4em}
    d)
    \hspace{2em}
    \begin{fmfgraph*}(70,70)
      \fmfstraight
      \fmftop{i2,o4}
      \fmfleft{i1,b2,b3,i2}
      \fmfright{o1,o2,o3}
      \fmfbottom{i1,o1}
      \fmf{fermion}{i2,v1}
      \fmflabel{$\nu$}{i2}
      \fmf{fermion}{v1,o4}
      \fmflabel{$\nu$}{o3}
      \fmf{dbl_plain_arrow,tension=1.5}{i1,v2}
      \fmflabel{A}{i1}
      \fmf{photon,label=$Z^0$,label.dist=-0.65cm}{v1,v2}
      \fmf{photon}{v2,o2}
      \fmflabel{$\gamma$}{o2}
      \fmf{dbl_plain_arrow}{v2,o1}
      \fmflabel{A}{o1}
    \end{fmfgraph*}
    \vspace{2em}
  \end{fmffile} 
  \caption{\label{fig:cartoon}
  Example diagrams of the NC1$\ga$ processes, including (a) a radiative process, (b) a baryonic resonance process, (c) an anomaly-mediated process, and (d) a coherent process. A single photon ($\ga$) is emitted in all of these NC neutrino ($\nu$)-nucleon ($N$) or neutrino-nucleus ($A$) interactions by exchanging Z-boson ($Z$).  Analyses can only measure the final state gamma ray, and cannot distinguish different primary processes of NC1$\ga$. Here, ``$N^*$'' represents a baryon resonance, and ``$M$'' stands for a neutral vector meson, such as an $\om$.}
\end{figure}
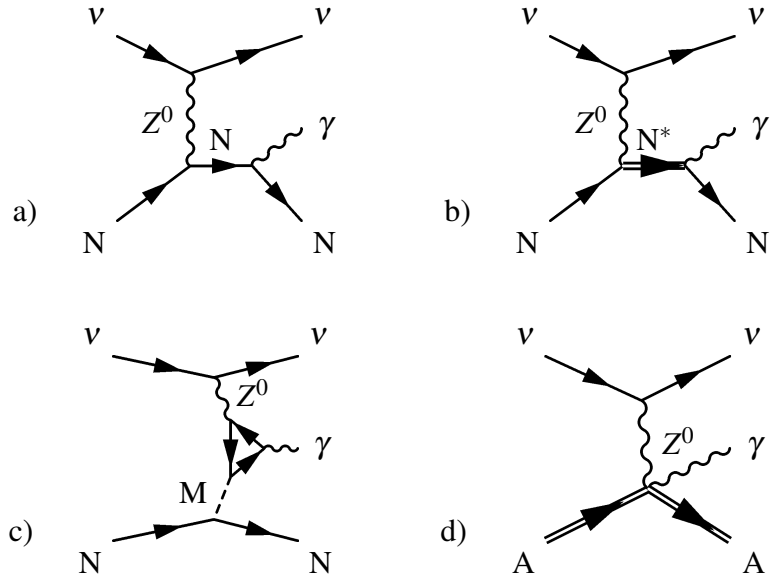

NC1$\ga$ is a rare process which has been identified as an important background process in $\nue$($\nuebar$) appearance oscillation experiments.  This process has significant theoretical uncertainties~\cite{Garvey:2014exa,Alvarez-Ruso:2014bla,Katori:2016yel,Alvarez-Ruso:2017oui}. A single photon with energy of order 100 MeV from NC1$\ga$ may be mistaken for the $\nue$($\nuebar$) appearance signal.  Figure~\ref{fig:cartoon} shows diagrams associated with the NC1$\ga$ process.  If the NC1$\ga$ process is related to a radiative decay of $\De$-resonance, a simple estimate of the cross-section based on a ratio of the branching ratios ($\De\to N\ga/\De\to N\pi$), gives the cross-section of NC1$\ga$ of $\sim 10^{-41}$~cm$^2$ per nucleon around the T2K beam energy.   

There was some interest in studying this process in the past~\cite{Gershtein:1980wu}, motivated by the low energy photon excess observed in the Gargamelle experiment~\cite{Alibran:1978jp}. One interpretation of the MiniBooNE excesses is NC1$\ga$ production. Recently, a series of new calculations of NC1$\ga$ have
been published. These models include contributions from previously ignored anomaly mediated photon production~\cite{Harvey:2007rd,Hill:2009ek}, a calculation based on the chiral effective field theory~\cite{Serot:2012rd,Zhang:2012aka,Zhang:2012xi}, a model including higher resonance contributions and nuclear media effect~\cite{Wang:2013wva}, and others~\cite{Efrosinin:2009zz,Jenkins:2009uq,Rosner:2015fwa}. These new calculations are consistent with the NC1$\ga$ background simulation used by MiniBooNE~\cite{Hill:2010zy,Zhang:2012xn,Wang:2014nat}. 
It has also been suggested that new physics processes could make NC1$\ga$-like final states which potentially explain MiniBooNE excesses, including heavy neutrino radiative decay  models~\cite{Gninenko:2009ks,Gninenko:2010pr,Dib:2011hc,Masip:2012ke,Duarte:2016miz} and massive neutral boson decay models~\cite{Ballett:2016opr,Bertuzzo:2018itn,Arguelles:2018mtc}. Some constraints on these models have been realized~\cite{Jordan:2018qiy}.  For Hyper-K, the NC1$\ga$ process is predicted to produce approximately 10\% of the background. However, given the 100\% theoretical uncertainties assigned to NC1$\ga$ in both neutrino and antineutrino modes, and the absence of measurements below 1~GeV, this is a source of systematic uncertainty that should be better understood.

This paper presents the result of the first search for NC1$\ga$ below 1~GeV in the T2K near detector data, which is relevant for current and future $\nue$($\nuebar$) appearance oscillation experiments. The NOMAD experiment at CERN performed the first search for NC1$\ga$, and set a limit on the total cross-section ratio of NC1$\ga$ to CC inclusive cross-section of $\NOMADratioA$ (90\% C.L.), at an averaged beam energy of $\EnuAve\sim\NOMADEnu$~GeV~\cite{Kullenberg:2011rd}. As discussed in this paper, the selection of NC1$\ga$  candidates is challenging for lower energy neutrino beams, and this measurement is of value to future experiments (Hyper-K and DUNE) in this energy range which rely on counting electron (anti)neutrinos in their detectors.

\section{T2K experiment}

\begin{figure}[tb]
  \includegraphics[width=0.5\textheight]{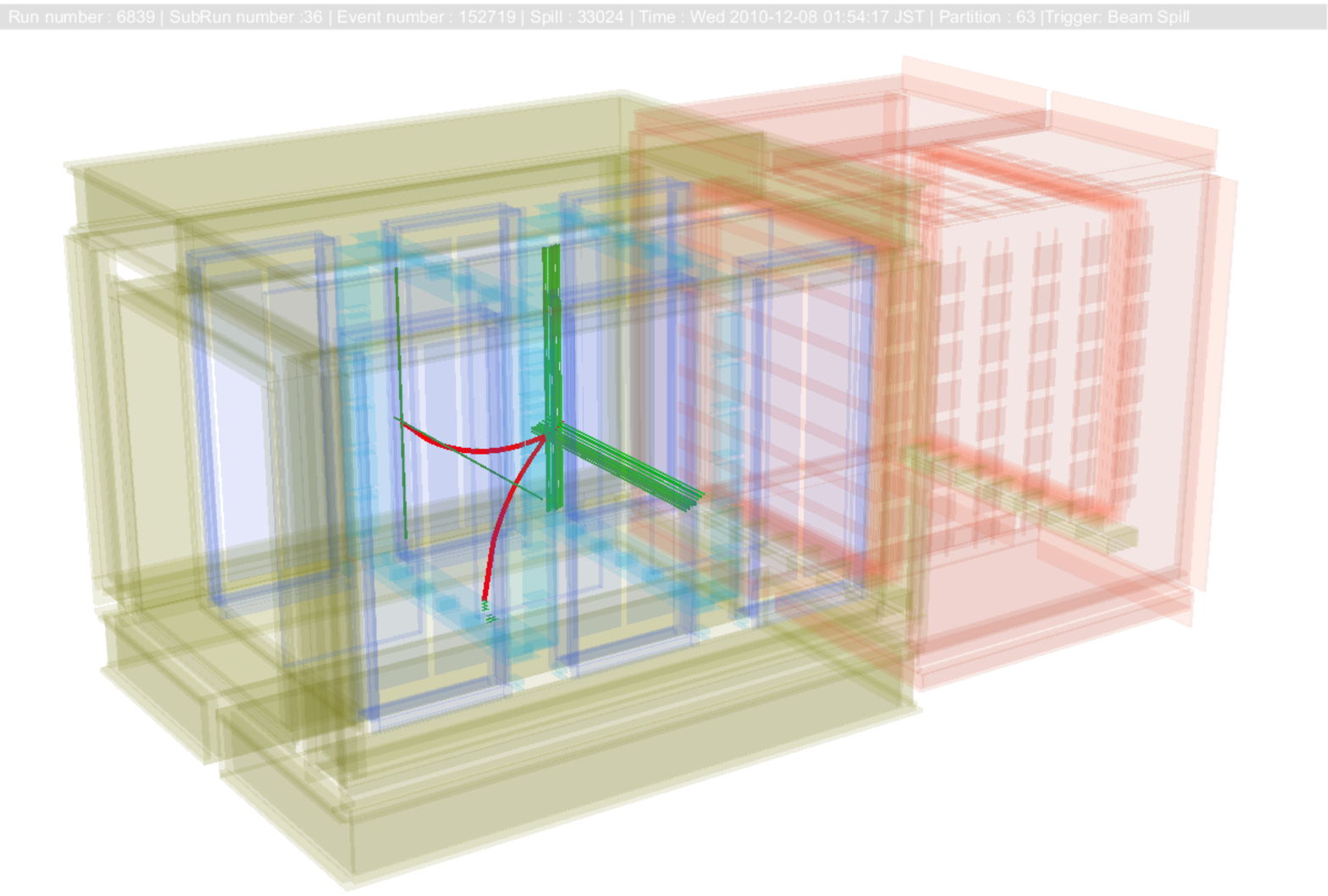}

  \caption{\label{fig:display}
    An example event display of a NC1$\ga$ candidate event from the data.
    The neutrino beam comes from the right. A neutrino interaction happens in FGD1 (light blue rectangular box) where green bars represent scintillation bars registered hits. Red tracks are reconstructed positive and negative electron-like tracks identified in TPC2 with opposite curvature due to the magnetic field. One particle reaches to the FGD2 to make another hits, and other track reaches to a surrounding sub-detector to leave hits. 
  }
  \end{figure}

T2K is a long-baseline neutrino oscillation experiment in Japan. Neutrinos are sent to the 50~kton Super-K detector with a baseline of 295~km. Primary protons are extracted from the 30~GeV J-PARC proton synchrotron to the dedicated neutrino beamline, where protons collide with a carbon target to produce secondary mesons, mainly pions. These mesons decay in the $\dpipe$~m long decay pipe to produce a tertiary neutrino beam. Depending on the current polarity of the magnetic focusing horns, the beamline can produce either $\numu$-dominant $\nu$-mode beam or $\numubar$-dominant $\nubar$-mode beam. The neutrino beam simulation incorporates hadron production data from NA61/SHINE experiment at CERN~\cite{Abgrall:2015hmv}. This analysis uses the $\nu$-mode beam data from November 2010 to May 2013, resulting total statistics of $\POT$ protons-on-targets (POTs). The details of the T2K neutrino beam line is described elsewhere~\cite{Abe:2012av}.

There are two near detectors, both located at baseline of 280~m, the on-axis near detector INGRID~\cite{Abe:2011xv} and off-axis near detector ND280. ND280 is a tracker detector which consists of several sub-detectors, including plastic scintillator tracker with radiator $\piz$-detector (P0D)~\cite{Assylbekov:2011sh}, plastic scintillator tracker fine-grained detectors (FGDs)~\cite{Amaudruz:2012esa}, gas time projection chambers (TPC)~\cite{Abgrall:2010hi}, electromagnetic calorimeters (ECal)~\cite{Allan:2013ofa}, and a side muon range detector (SMRD)~\cite{Aoki:2012mf}. The sub-detectors, going downstream in the neutrino flux are the P0D, followed by two FGDs and three TPCs which alternate to make the tracking region of ND280. The P0D and the FGDs provide target mass and vertex measurements, and the subsequent TPCs provide tracking measurements. All sub-detectors are immersed in a $\magnet$~T dipole magnetic field, and track measurements in the TPCs provide charge and momentum measurements of charged particles. Fig.~\ref{fig:display} is an event display of a NC1$\ga$ candidate event. The neutrino interaction is identified in the first FGD (FGD1), and subsequent TPC2 measures electron and positron tracks. A rectangular region $\FidX$~cm (x)$\times\FidY$~cm (y)$\times\FidZ$~cm (z) in the FGD1 is defined to be the fiducial volume of this analysis where the z-axis is the direction of the beam, the y-axis points upward, and the x-axis is chosen to complete the right-hand Cartesian coordinate. The target material is polystyrene CH$_2$ and the number of target nucleons is $\target$ with $\eFGDmass$\% error.

The detector Monte Carlo (MC) simulation is based on GEANT4~\cite{GEANT4}, and neutrino interactions are simulated by the NEUT event generator version  $\vNEUT$~\cite{Abe:2018wpn}. Note, the normalization of the NC1$\ga$ model used in NEUT v.$\vNEUT$ was found to be roughly 50\% lower than more recent calculations~\cite{Wang:2015ivq,Lasorak:2016acm}, however, it does not affect our analysis result. 

\section{Event selection}

\begin{figure}[t]
  \includegraphics[width=0.5\textheight]{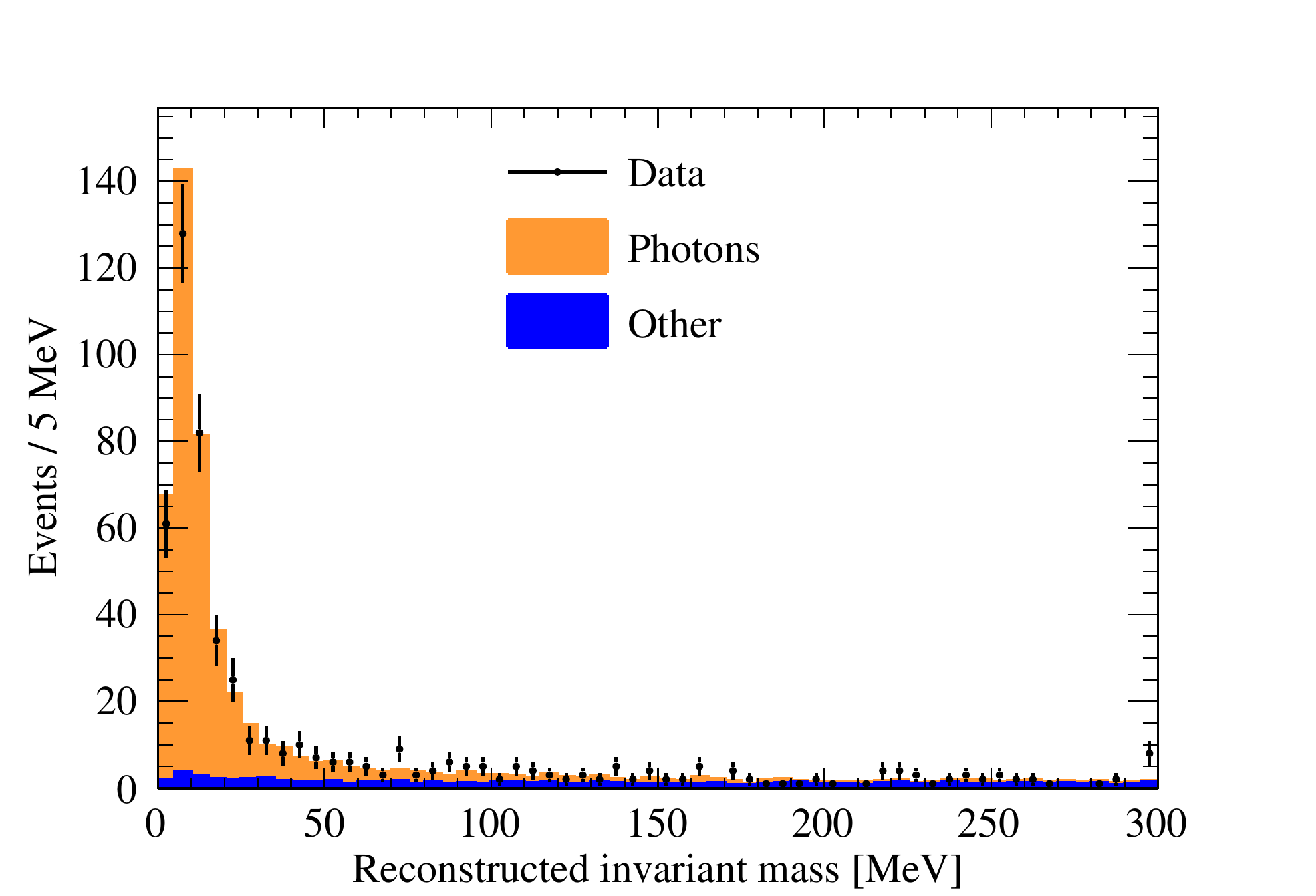}
  \caption{\label{fig:inv}
    Invariant mass distribution of the photon sample. To select photons we apply a cut on the reconstructed invariant mass, $M_{inv}<50$~MeV. 
  }
  \end{figure}

The event selection of the photon sample has been developed for the $\nue$ charged current ($\nue$CC) measurements in ND280~\cite{Abe:2014usb,Abe:2014agb,Abe:2014nuo}, where photons make a major background for $\nue$($\nuebar$) analysis. Thus, for these analyses the photon sample was made to study the background distribution. In this analysis, instead, we use this sample to search for NC1$\ga$. Photons are identified from two tracks. These tracks are required to have opposite charges. Tracks should start from within the fiducial volume of FGD1 scintillator tracker. To maintain the quality of momentum reconstruction, these tracks should leave at least 18 reconstructed clusters in TPC2 corresponding to a $\sim$18~cm track if they are straight in the direction of the beam. Particle identification (PID) based on energy loss measured in TPC is applied to select electron-like or positron-like tracks. The starting points of these tracks have to be within 10~cm of each other. Then, the invariant mass ($M_{inv}$) is reconstructed from the measured momenta of two electron-like tracks with opposite charges. Fig.~\ref{fig:inv} shows the invariant mass distribution. As can be seen, low invariant mass is dominated by photons and we choose $M_{inv}<50$~MeV to construct the photon sample. The photon purity in the sample reaches $\purity$\%, however, the majority of the photons are generated outside of the fiducial volume. 

\begin{figure}[t]
  \includegraphics[width=0.35\textheight]{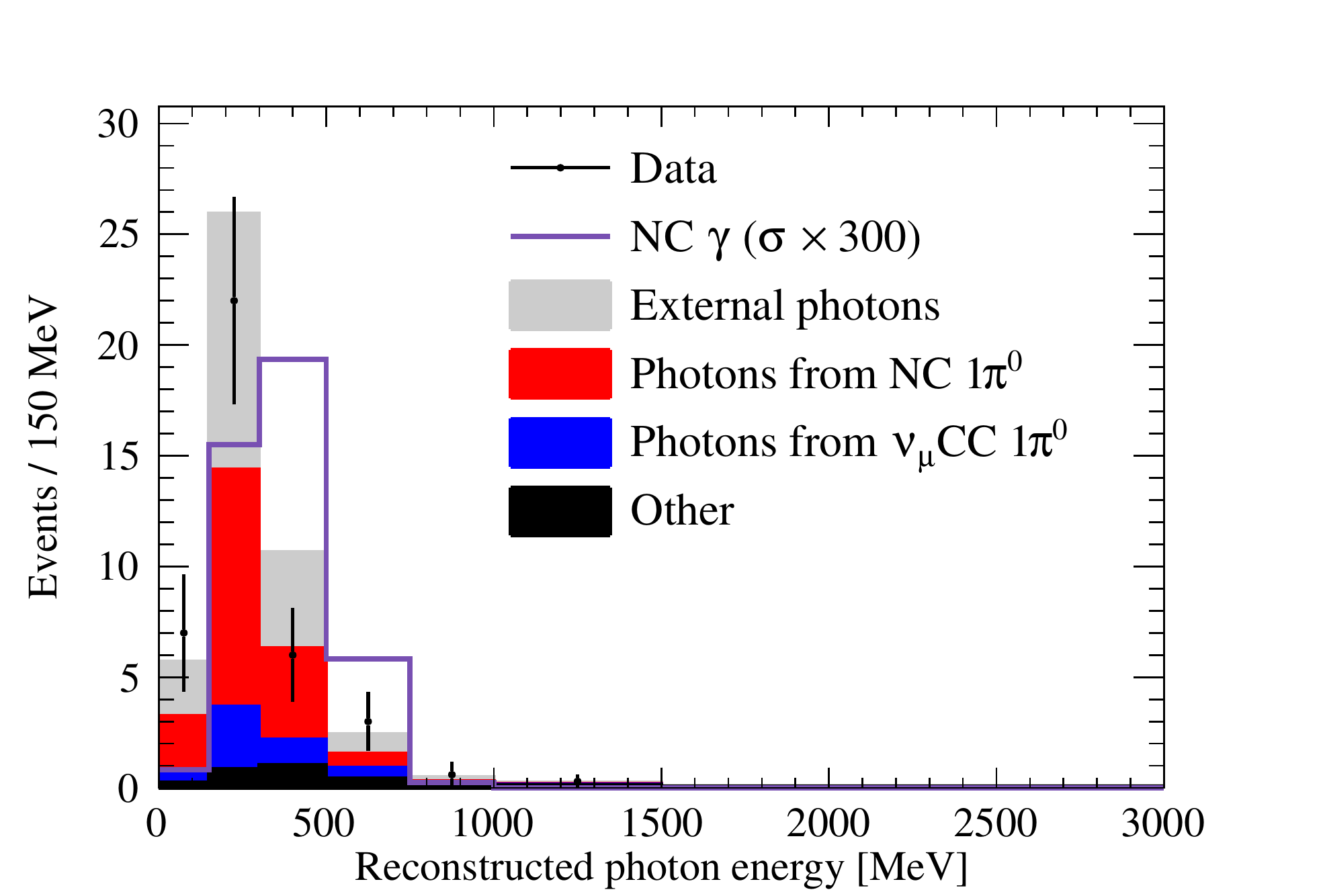}
  \includegraphics[width=0.35\textheight]{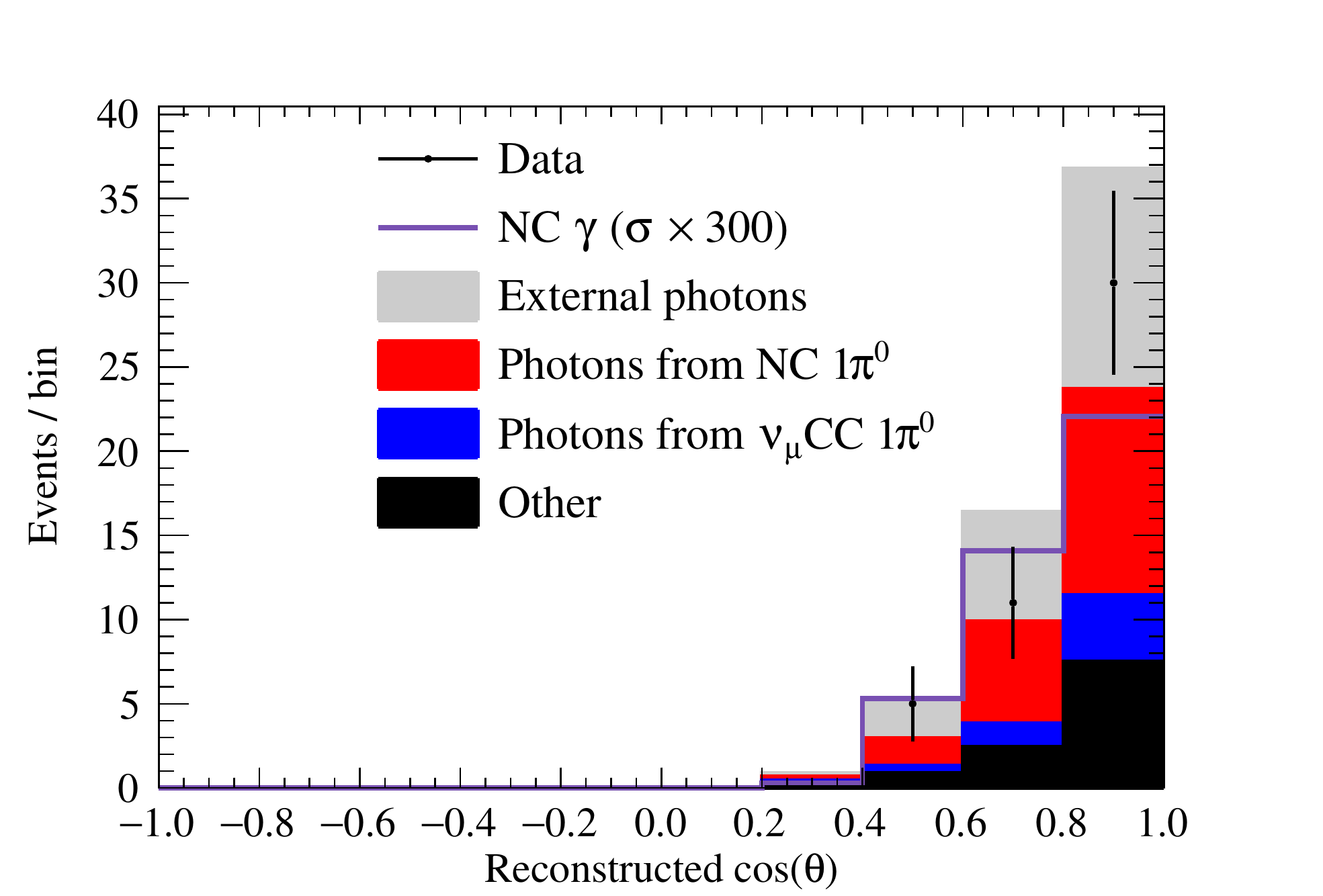}
  \caption{\label{fig:selection}
    Reconstructed energy and scattering angle distribution of the NC1$\ga$ sample. The data is shown with markers, and the simulation is shown as a histogram. The simulation is stacked with different primary processes to produce photons. Note, the NEUT NC1$\ga$ prediction is scaled up by a factor 300 to be visible. 
  }
  \end{figure}

We further use the surrounding sub-detectors to remove photons which are not within the fiducial volume to make a NC1$\ga$ sample. First, we remove any events associated with muons detected in any TPC. These interactions are most likely CC interactions and they are backgrounds of this analysis. Second, we remove events with reconstructed clusters in the surrounding ECals and P0D that are not associated with the gamma, because NC interactions in these sub-detectors may produce photons which convert in the FGD1 detector mimicking photons generated in the FGD1 detector. After these cuts, we selected $\Nevt$ events to construct the NC1$\ga$ sample. Fig.~\ref{fig:selection} shows the reconstructed energy and scattering angle distributions of the NC1$\ga$ sample. The peak of photon momentum is around 200~MeV/c and peaked in the forward direction. According to our simulation, the selection efficiency for NC1$\ga$ events is $\efficiency$\%. However, the sample is dominated by internal or external backgrounds. Internal backgrounds are mainly single photons from asymmetric decays of $\piz$s produced by NC interactions in the fiducial volume. External backgrounds are photons generated from outside of the fiducial volume leaving no traces in sub-detectors, and are converted to $e^+-e^-$ pairs in the FGD1 fiducial volume. Because of the presence of these backgrounds, the expected signal fraction, {\it i.e.}, the fraction of photons produced by NC1$\ga$ process in the FGD1 fiducial volume is less than 1\% according to our simulation. Based on uncertainties in the background processes, we could not detect the NC1$\ga$ process from this analysis, and the remaining part of this paper focuses on setting a cross-section limit on this process.

\section{Systematic errors}

\begin{table}[tb]
      \begin{tabular}{c | c}
        Error type & values (\%)  \\
        \hline \hline
        Statistical error  & $\pm\psta$ \\
        \hline                                               
        pion background    & $+\psyp$/$-\nsyp$ \\
        External background& $+\psyb$/$-\nsyb$ \\
        Flux         & $\pm\psyf$ \\
        Detector     & $\pm\psyd$ \\
        Neutrino interaction  & $+\psyx$/$-\nsyx$ \\
        \hline                                              
        Total error  & $+\ptot$/$-\ntot$ \\        
        \hline \hline
    \end{tabular}
    
    \caption{\label{tab:error}
      The summary table of the errors on this analysis. The largest source of uncertainty comes from the asymmetric uncertainty on the external background.}    
\end{table}

The NC1$\ga$ sample is dominated by internal and external backgrounds. Thus, the NC1$\ga$ cross-section measurement is limited by these backgrounds. To constrain the internal background, we use NC$\piz$ data from MiniBooNE~\cite{AguilarArevalo:2009ww} in the NUISANCE framework~\cite{Stowell:2016jfr} to estimate errors associated the $\piz$ production. Uncertainties were set on parameters of the $\piz$ production model to cover the shape and normalization differences between the model predictions and the MiniBooNE data. This gives around 15\% systematic error on the prediction of the NC1$\piz$ rate~\cite{Lasorak:2018odg}. The details of the evaluation of this systematic uncertainty are given in \ref{sec:A1}.

To constrain the external background, we estimate the variations of the mass distribution outside of the fiducial volume, and photon propagation from external materials. We use CC inclusive data sample collected from the outside layers of FGD1, which is dominated by muons produced by neutrino interaction with materials surrounding the fiducial volume. The data-MC disagreement is around $\CCincla$\% except for up-going events where the disagreement is $\CCinclb$\%. These data suggests that the up-going external background is not properly modeled, and this would add an additional systematic error to the up-going photon external background. Thus, we limit our measured region to be $0^\circ<\ph<\PhiL$ and $\PhiH<\ph<360^\circ$, where $\ph$ is the angle of reconstructed photon direction projected on a x-y plane with $\ph=0$ on the +x axis. By removing up-going events in the sample, $\NevtPhi$ events are left in the NC1$\ga$ sample.

Through the MC we evaluate the material and density errors affecting the photon propagation from inactive materials to the fiducial volume. For this, we define the photon effective mean free path (EMFP) $\EMFP$ to find the uncertainty of the external photon backgrounds which produce the $e^+-e^-$ pairs. We estimate the variations of EMFP in the dead materials by using mass error from technical reports and from surrounding muons measurement. We then propagate changes on the EMFP to the probability that a photon arrives in the FGD1. Although this is a reasonable approach to estimate the variation of the external photon background from the simulation, this method creates large variation in the number of external photon events, {\it i.e.}, small errors in density and material composition at the production and propagation result in a large variation in the conversion points in FGD1 after propagation through dead materials. We estimate a 27\% systematic uncertainty on the prediction of the external background. This procedure is described in \ref{sec:A2}.

After evaluating the errors associated to backgrounds, we also add uncertainties coming from simulation of neutrino flux, detector, and other neutrino interaction processes. Table~\ref{tab:error} is the summary of all errors for this analysis. The largest error is the external background variation which is the limiting systematic uncertainty in this analysis. Internal background, mainly errors associated to pion production, and statistics also contribute to the final error of this analysis.

\section{Result}

\begin{figure}[tb]
  \includegraphics[width=0.6\textheight]{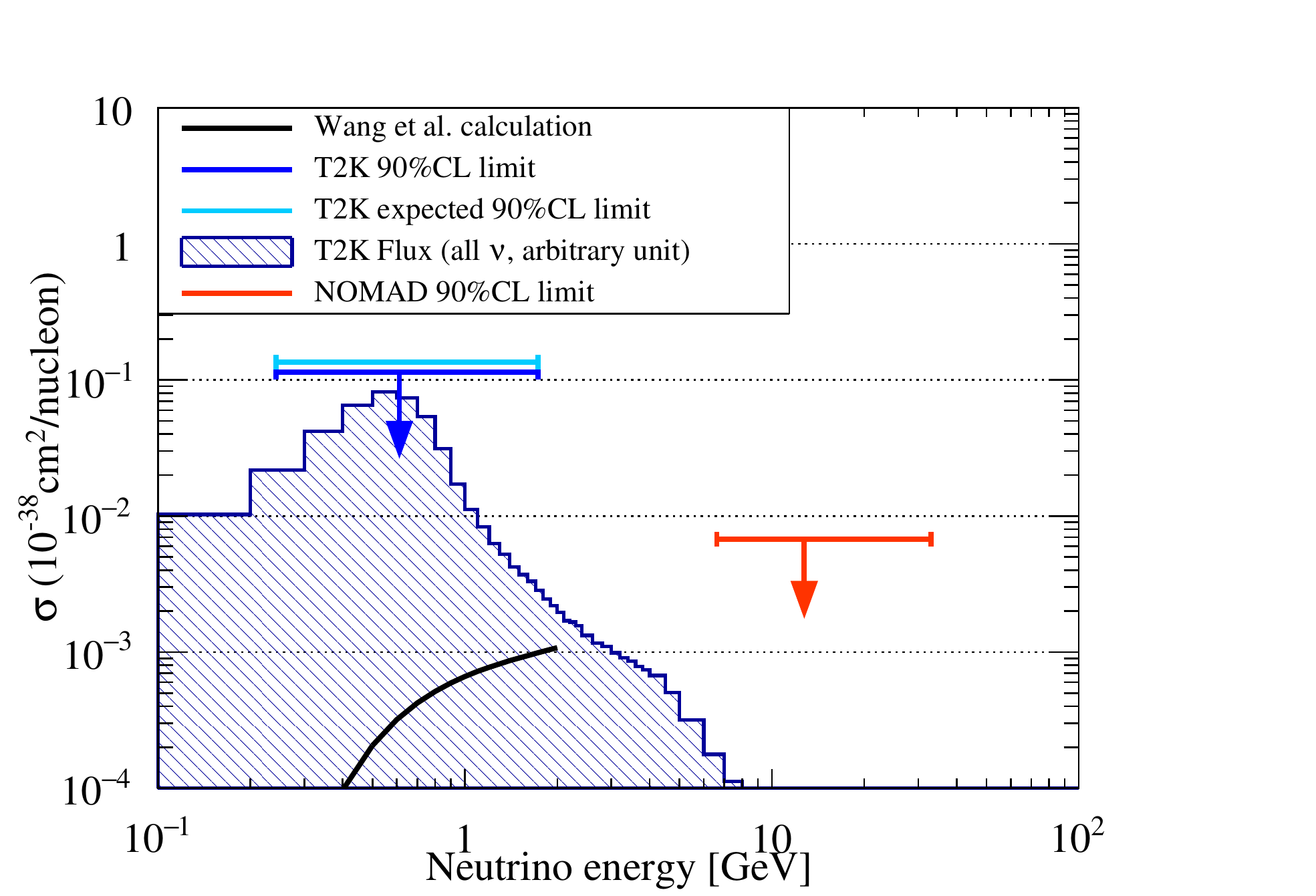}
  \caption{\label{fig:limit}
    The NC1$\ga$ cross-section limits from this analysis. The cyan line is the 90\% C.L. sensitivity from the MC, and the blue line shows the 90\% C.L. limit from this analysis. Both are averaged over J-PARC neutrino flux. The blue histogram shows the distribution (arbitrary unit) of the J-PARC $\nu$-mode muon neutrino flux used by this analysis~\cite{Abe:2012av}. The results are compared with one of recent calculation (black curve)~\cite{Wang:2013wva}. Note, the model is terminated at the neutrino energy $=2.0$~GeV. The result is also compared with the results from NOMAD (red line)~\cite{Wu:2007ab,Kullenberg:2011rd}.
  }
  \end{figure}

After evaluating all errors, we generate sets of the background simulations (toy MC), and from this distribution and data, we set the limit of the expected number of events from the NC1$\ga$ process. By using the MC, we convert this limit to the total NC1$\ga$ cross-section limit. Thus, our result is a model dependent cross-section limit. The total cross-section limit derived by this method is found to be $<\limA$~cm$^2$ (90\% C. L.).

Fig.~\ref{fig:limit} shows the result. Cyan and blue lines represent the sensitivity and the limit from this analysis, and the blue histogram shows the flux shape used by this analysis. The black curve is a recent calculation of the NC1$\ga$ cross-section~\cite{Wang:2013wva}. As can be seen, our limit is far from the expected signal. This is mostly due to uncertainties of internal and external background predictions where we rely on external data and simulation to evaluate them instead of constrain them by {\it in situ} measurements. Nevertheless, we achieve to set the first limit on this process below 1~GeV. The results are also compared with those from NOMAD~\cite{Kullenberg:2011rd}. NOMAD performed a search for $NC1\ga$, 
and NOMAD reported the upper limit of the process in terms of the cross-section ratio to CC inclusive cross-section, $\NOMADratioA$ (90\% C. L.). By multiplying this ratio with the NOMAD reported CC inclusive cross section~\cite{Wu:2007ab}, we calculate the $NC1\ga$ total cross-section upper limit from NOMAD, $<\NOMADlimA$~cm$^2$ (90\% C. L.)~\cite{Mishra}.

\section{Outlook}

In this article, we described the search for NC1$\ga$ process below 1~GeV, using the T2K off-axis near detector. Although we found $\NevtPhi$ NC1$\ga$ candidate events, these events are consistent with predicted background events and we set the first limit on the NC1$\ga$ cross-section below 1~GeV, at $<\limA$~cm$^2$ (90\% C.L.). An excellent tracking system allows to construct a $\purity$\% pure photon sample, however, there are two main factors which limit our analysis. First, the analysis does not use an internal constraint on NC$\piz$ production rate, and we rely on external data to understand NC$\piz$ production rate uncertainties. Ideally, we should utilize a simultaneous measurement of photons and $\piz$s so that the systematics of $\piz$ production rate can be constrained. NC$\piz$ production has been measured in P0D~\cite{Abe:2017urf}, and such measurement in FGD has been developed~\cite{LeonPickard}. Second, an internal constraint for external background is not available, and we rely on mainly simulation to estimate the incoming photon background. Such background could be internally measured if the detector had a large active veto region, and similarly could be suppressed if the detector had less dead material between the active veto and the fiducial volume. This may be achieved by the P0D where larger fiducial volume than FGDs can reduce external background. 
New active detectors developed for T2K, such as WAGASCI~\cite{Kin:2017way} may overcome these problems and set a better limit by utilizing better tracking with relatively larger fiducial volume. Some current neutrino experiments, such as MINERvA~\cite{Wolcott:2015hda}, MicroBooNE~\cite{Acciarri:2016smi}, SBND and ICARUS~\cite{Antonello:2015lea} have larger fiducial volumes with less inactive detector regions, and these experiments have better control for both internal and external backgrounds, and they also have the chance to make the first measurement of NC1$\ga$ process. 


\section*{Acknowledgment}
We thank the J-PARC staff for superb accelerator performance. We thank the CERN NA61/SHINE Collaboration for providing valuable particle production data. We acknowledge the support of MEXT, Japan; NSERC (Grant No. SAPPJ-2014-00031), NRC and CFI, Canada; CEA and CNRS/IN2P3, France; DFG, Germany; INFN, Italy; National Science Centre (NCN) and Ministry of Science and Higher Education, Poland; RSF, RFBR, and MES, Russia; MINECO and ERDF funds, Spain; SNSF and SERI, Switzerland; STFC, UK; and DOE, USA. We also thank CERN for the UA1/NOMAD magnet, DESY for the HERA-B magnet mover system, NII for SINET4, the WestGrid and SciNet consortia in Compute Canada, and GridPP in the United Kingdom. In addition, participation of individual researchers and institutions has been further supported by funds from ERC (FP7), "la Caixa” Foundation (ID 100010434, fellowship code LCF/BQ/IN17/11620050), the European Union’s Horizon 2020 Research and Innovation programme under the Marie Sk\l{}odowska-Curie grant agreement no. 713673 and H2020 Grant No. RISE-GA644294-JENNIFER 2020; JSPS, Japan; Royal Society, UK; the Alfred P. Sloan Foundation and the DOE Early Career program, USA.

\appendix
\section{Background error estimation}

\subsection{Internal background error estimation\label{sec:A1}}

In this section, we discuss the error estimation of the largest internal background, NC$\piz$ production rate. There is a tension in $\piz$ momentum space between the NEUT prediction and the MiniBooNE NC$\piz$ data. Six parameters are used to cover this discrepancy. First, uncertainties are set on the $\piz$ production model parameters, including the resonant axial mass ($M_A^{RES}=0.95\pm 0.15$ GeV), the $C_5^A$ form factor normalization ($C_A^5=1.01\pm 0.12$), and the isoscalar contribution normalization ($I_{1/2}=1.30\pm 0.20$). Second, three additional {\it ad hoc} systematic parameters are added. The first one is the shift of the $\De$ resonance peak and we introduced a 0.4\% systematic error. The second one is the width of the $\De$ resonance and we introduce a 14\% systematic error. And the third is the normalization of NC coherent $\piz$ production channel and we introduce a 100\% error. These six systematic errors cover the difference between MiniBooNE NC$\piz$ data and NEUT. The resulting systematic uncertainties used in this analysis are presented in the second row of Table~\ref{tab:error}.

\subsection{External background error estimation\label{sec:A2}}

In this section, we discuss the error estimation of the largest external background, the external photon conversion rate in the fiducial volume. Although changing parameters in the simulation allows us to evaluate the error in the number of photons arriving in the fiducial volume from external materials, this is CPU intensive, and impractical.  Instead, we define $\EMFP$, the effective mean free path of the photon in the material, and we apply changes to $\EMFP$ to evaluate photon background systematics.

The number of photons $N(x_i)$ radiated at a given position $x_i$ away from its creation point $x^0_i$ is written in the following way,
\beq
N(x_i)=N\left(x^0_i\right)\exp\left[-|x_i-x^0_i|/\EMFP\right]~.
\label{eq:mfp}
\eeq
For external photons converted in the fiducial volume, we generate the systematic variation in the simulation by applying a weight defined by the ratio of the Eq.~\ref{eq:mfp}:
\beq
w\equiv\frac{\exp\left[-|x_i-x^0_i|/\EMFP'\right]}{\exp\left[-|x_i-x^0_i|/\EMFP\right]}~.
\label{eq:w}
\eeq
To apply this weight to a simulated event, one must know the distance that the photon traverses in the particular material, $|x_i-x^0_i|$ and $\EMFP$. To calculate the nominal $\EMFP$ (in the numerator of Eq.~\ref{eq:w}), we proceed in two ways: First, for photons with starting points within the dead materials, $\EMFP$ is found by fitting to the MC sample for different locations surrounding the FGD.  Second, for photons starting in active regions of the sub-detectors, we calculate it analytically, using 
\beq
1/\EMFP=\sum_i m_i/\la^i~.
\eeq
Here, $m_i$ and $\la_i$ are the mass fraction and the mean free path of a material $i$. The mean free path $\la_i$ of photons for arbitrary material $i$ can be written as~\cite{Tanabashi:2018oca} 
\beq
1/\la=3.1\al r_e^2D_{atom}\left[Z^2(L_{rad}-f(z))+ZL_{rad}'\right]~,
\label{eq:Tsai}
\eeq
where we use the fine structure constant $\al$, classical electron radius $r_e$, atomic density $D_{atom}$, atomic number $Z$, Tsai's radiation length $L_{rad}$ and $L_{rad}'$ with a high order correction $f(Z)$ ~\cite{Tsai:1973py}.
Using Eq.~\ref{eq:Tsai}, one can modify the density (hence the mass) and the composition of the material to change the mean free path. The mass variations of materials, where a typical systematic uncertainty is order few \%, are derived from detector design reports~\cite{Assylbekov:2011sh,Amaudruz:2012esa,Abgrall:2010hi,Allan:2013ofa,Aoki:2012mf} and CC inclusive data leaving signals at the outer layers of the fiducial volume of FGD1. 

We apply this for photons coming from all directions, traverse different sub-detectors and materials, to find the distribution of external background variations. The shape of the weights for photons traversing a large distance of material is skewed towards high number of events. The skew reflects that a small change in the density of materials causes large errors in the number of photons converted in the fiducial volume. Note that the asymmetry of the errors was taken into account for the analysis and this is shown in the third row of Table~\ref{tab:error}.

\bibliography{ncgamma}

\end{document}